\documentclass[12pt]{iopart}

\usepackage{graphicx}
\begin{document}

\title[]{High momentum lepton pairs from jet-plasma interactions}

\author{Simon Turbide and Charles Gale}

\address{Department of Physics, McGill University, 3600 University Street, Montreal, Canada H3A 2T8}

\begin{abstract}
We discuss the emission of high momentum lepton pairs ($p_T>4$ GeV) with low invariant masses ($M\ll p_T$) in central 
Au+Au collisions at RHIC ($\sqrt{s_{NN}}=$200 GeV).  The spectra of dileptons produced through
interactions of quark and antiquark jets with the quark-gluon plasma (QGP) have been calculated.  Annihilation and Compton scattering processes, as well as processes benefitting from collinear enhancement, including Landau-Pomeranchuk-Migdal (LPM) effects, are calculated and convolved with a one dimensional hydrodynamic expansion. The jet-induced contributions are compared to thermal dilepton emission and Drell-Yan processes, and are found to dominate around $p_T=$4 GeV.
 
\end{abstract}


%\maketitle

\section{Introduction}

The search for the creation of a quark-gluon plasma (QGP) in relativistic heavy ion collisions has driven several experiments and produced many theoretical calculations. In this context, real and virtual photons can be used to probe the high temperature and density phase in these collisions: owing to a large mean free path, electromagnetic radiation suffers only little final state interactions~\cite{Feinb:76}.  Moreover, recent calculations~\cite{prlphoton,TGJM:05} suggest that the direct interaction of jets with the medium could constitute an important source of real photons.  Considering that the high-$p_T$ suppression observed at RHIC in hadron spectra~\cite{phenix2} can be explained by the quenching of jets as they go though the plasma, the production of electromagnetic radiation by jet-plasma interactions is but another manifestation of the same phenomenon. In this work we examine the virtual photons produced by the passage of jets in the QGP, and then decay into lepton pairs.  The dileptons produced by jet in-medium bremsstrahlung enjoy an enhancement related to collinear singularities, and were missing in Ref.~\cite{Turbide:2006mc}. They are now calculated, providing a complete calculations of in-medium dileptons from jets, up to the next-to-leading order in $g_s$. We finally compute and compare the dilepton sources associated with jets and perturbative QCD, at RHIC energies.

\section{Dilepton production}

\begin{figure}
  \begin{center}
  \includegraphics[width=0.6\textwidth]{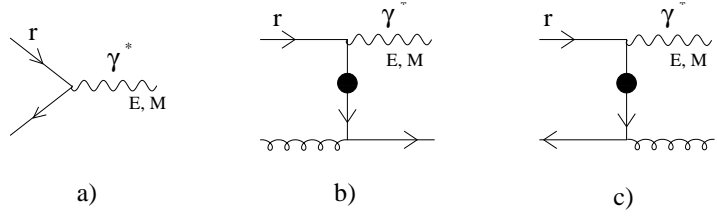}
  \end{center}
  \caption{\label{direct} Physical processes without a collinear enhancement. }
\end{figure}

From finite-temperature field theory~\cite{vmd91}, the production rate of a lepton pair with momentum $p$, invariant mass $M$ and energy $E$ is
\begin{equation}
\frac{d^4 R^{e^+e^-}}{d^4p}=\frac{2\alpha}{3\pi M^2}E\frac{d^3 R^{\gamma^*}}{d^3p}=\frac{2\alpha}{3\pi M^2}\frac{1}{(2\pi)^3}\frac{\mbox{Im}\Pi_\mu^{{\rm R}\,\mu}}{1-e^{E/T}}\,,
\end{equation}
where $\Pi_\mu^{{\rm R}\,\mu}$ is the retarded photon self-energy. The real and virtual photons are related by $E\,d^3 R^{\gamma}/d^3p=\lim_{M\to 0}E\,d^3 R^{\gamma^*}/d^3p$. In the hard thermal loop (HTL) resummation formalism~\cite{htl}, the annihilation and Compton scattering diagrams from Fig.~\ref{direct} are evaluated.  The solid circles in this figure indicate resummed propagators.  In this framework, all distributions are thermal.  Using relativistic kinetic theory, the thermal distribution of the incoming particle $n_{\rm FD}(r)$ is isolated, and substituded by a jet distribution $f_{jet}(r)=d^6 N_{jet}/d^3xd^3r$ (see Ref.~\cite{Turbide:2006mc} for more details).  The time evolution of the space-time distribution of jets is obtained with the help of the Arnold-Moore-Yaffe (AMY) formalism~\cite{AMY}, which include the gluon bremsstrahlung energy loss, by treating correctly the Laudau-Pomeranchuk-Migdal (LPM) effect [up to ${\cal O}(g_s)$ corrections].  We obtain the non-collinear dilepton yield by convolving the production rate with the hydro expansion
\begin{equation}
\label{yield_noncoll}
\frac{d^4 N^{e^+e^-}_{non-coll}}{dM^2d^2p_T dy}=\int d\tau\,\tau\int d^2x_\perp\int dp_z\frac{1}{2\sqrt{M^2+p_T^2+p_z^2}}\frac{d^4 R^{e^+e^-}}{d^4p}\,,
\end{equation}
where $\tau=\sqrt{t^2-z^2}$ is the proper time.  We assume a purely longitudinal expansion of the fireball~\cite{Bjorken}, such that at each point the temperature evolves according to $T({\bf r}_\perp,\tau)=(\tau_i/\tau)^{1/3}T({\bf r}_\perp,\tau_i)$.

\begin{figure}
  \begin{center}
  \includegraphics[width=0.6\textwidth]{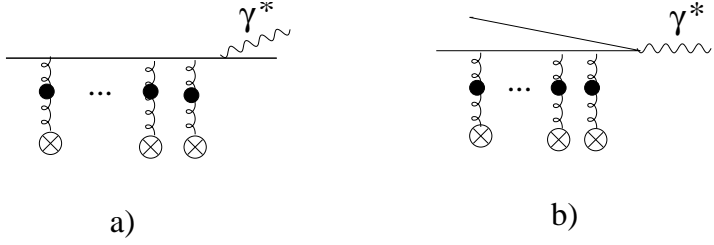}
  \end{center}
  \caption{\label{brem}  Bremsstrahlung and annihilation processes with LPM effect. }
\end{figure}

The dilepton production rate for processes that benefit from a collinear enhancement (see Fig.~\ref{brem}) with LPM effects, i.e. including an infinite sum of diagrams with different number of scatterings with soft gluons, has been extended from real photons~\cite{AMY} to virtual photons in Ref.~\cite{AGMZ}. Since the collinearity between the incoming jet and the emitted photon has been assumed in that work, the resulting $\Pi_\mu^{R\, \mu}$ is meant to be restricted to the region $M\ll p_T$. For an explicitly thermal fermion distribution, $n_{\rm FD}$, we may write
\begin{equation}
\frac{d^4 R^{e^+e^-}_{coll}}{d^4p}=24 \int_0^\infty dr\, \frac{r^2\, E}{(2\pi)^3 p^2}\, n_{\rm FD}(r) \frac{d^3 \Gamma}{dp dM^2 dt} \left(1-\frac{\theta(p-r)}{2}  \right)\,,
\end{equation}
where $d\Gamma/dt$ is the quark to dilepton transition rate. The $\theta(p-r)$ is included to avoid double counting in the annihilation process, since $n_{\rm FD}^q(r)n_{\rm FD}^{\bar{q}}(p-r)=n_{\rm FD}^q(p-r)n_{\rm FD}^{\bar{q}}(r)$.  After subtracting from $d\Gamma$ the leading order annihilation (Fig.~\ref{direct}a), again to avoid double counting, the midrapidity dilepton yield from the collinear processes induced by jets is

\begin{equation}
\label{yield_coll}
\frac{d^4 N^{e^+e^-}_{coll}}{dM^2d^2p_T dy}=\int dt \int dr_T\, \frac{r_T}{p_T}\, \frac{d^3 N^{q\bar{q}}_{jet}}{d^2r_T dy}\,\frac{d^3 \Gamma}{dp_T dM^2 dt}\,.
\end{equation}
The fluctuations in the creation position of jets (and the resulting different path lengths in the medium) are included in our calculation. 
%both Eqs.~\ref{yield_noncoll} and ~\ref{yield_coll}.

\section{Results}

\begin{figure}
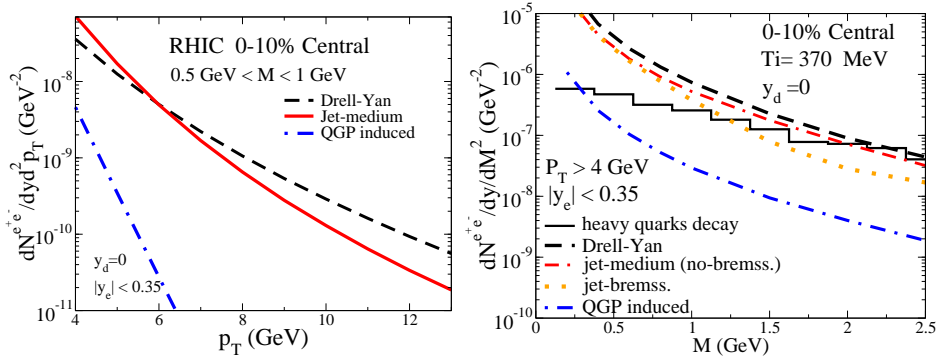

  \begin{center}
  \includegraphics[width=0.38\textwidth]{dilep_pt_05_1GeV.eps}
 \includegraphics[width=0.4\textwidth]{dilep_lowm.eps}
\end{center}

\caption{\label{dilep} (Color online) Momentum (left) and invariant mass (right) distribution of midrapidity dileptons in central Au+Au collisions at RHIC.  Left panel: solid line, jet-medium contribution; dashed line, Drell-Yan; dot-dashed line, QGP induced radiation. Right panel: solid line, semileptonic decay of heavy quarks; double dash-dotted and dotted line respectively non-collinear and collinear jet-medium contributions.}
\end{figure}

Our results for the contribution of in-medium jet-induced dileptons are shown in Fig.~\ref{dilep}, for a QGP with initial temperature $T_i=$370 MeV, with fixed $\alpha_s=0.3$.  To highlight the importance of this contribution, we display also the QGP induced radiation, i.e. the dilepton yield when we don't substitute $n_{\rm FD}$ by $f_{jet}$, and the Drell-Yan processes~\cite{Turbide:2006mc}.  For the integrated mass window $0.5 < M < 1$ GeV (left panel), the jet-induced contribution (solid line) turns out to be dominant around $p_T$=4 GeV, and orders of magnitude above the QGP induced radiation (dot-dashed line), for $p_T>$ 4 GeV, in good analogy with what has been found for real photons in Ref.~\cite{TGJM:05}.   In the right panel, we show the invariant mass dilepton distribution, integrated over all $p_T$ above 4 GeV, but as the dilepton yields are steep functions of $p_T$, the $p_T$ integrations are peaked around $p_T=4$ GeV.  Here the total jet-induced contribution has been split into the collinear (dotted line) and non-collinear (double dash-dotted line) dileptons, and we can observe that the former contribution becomes less and less important with increasing $M$.  Both processes are however of the same size below $M=1$ GeV, where they are also both as important as the Drell-Yan contribution.  We also provide in the right panel an estimate of the heavy quark decay contribution~\cite{mangano}, whithout the inclusion of jet energy loss.  The lepton pair production from jets beiing as big as this upper limit background, we expect the experimental detection of a jet-plasma contribution to be feasible.

\section{Conclusion}

We have calculated for the first time, the in-medium jet induced dileptons up to the next-to-leading order in $g_s$, at high-$p_T$ and low-$M$. This has been accomplished by substituting the phase-space distribution of incoming thermal quarks by jets distribution. The collinear processes have been shown to be as important as the non-collinear processes below $M=1$ GeV.  The jet-plasma contributions turn out to be order of magnitude above the radiation induced by the QGP, and as important as Drell-Yan and the background coming from the decay of heavy quarks around $p_T=$ 4 GeV.  Further studies of dilepton production in more involved time-evolution models are currently in progress.

\ack
We thank F. G\'elis and G.D. Moore for helpful discussions. This work was supported in part by the Natural Sciences and Engineering Research Council of Canada.

\end{document}